\documentstyle[psfig]{mn}

\newif\ifAMStwofonts

\newcommand{\tbl}[1]{\mbox{Table \ref{tbl:#1}}}
\newcommand{\fig}[1]{\mbox{Figure \ref{fig:#1}}}
\newcommand{\eqn}[1]{\mbox{Equation (\ref{eqn:#1})}}
\newcommand{\sct}[1]{\mbox{Section \ref{sct:#1}}}
\newcommand{\app}[1]{\mbox{Appendix \ref{app:#1}}}
\newcommand{\gesim}{\,\raisebox{-0.4ex}{$\stackrel{>}{\scriptstyle\sim}$}\,}
\newcommand{\lesim}{\,\raisebox{-0.4ex}{$\stackrel{<}{\scriptstyle\sim}$}\,}

\def\bvec#1{{\mbox{\boldmath $#1$}}}
\def\hatalpha{\bvec{\hat{\alpha}}}
\def\Dos{D_{\mbox{{\small os}}}}
\def\Dod{D_{\mbox{{\small od}}}}
\def\Dds{D_{\mbox{{\small ds}}}}

\ifoldfss
  \ifCUPmtlplainloaded \else
    \NewTextAlphabet{textbfit} {cmbxti10} {}
    \NewTextAlphabet{textbfss} {cmssbx10} {}
    \NewMathAlphabet{mathbfit} {cmbxti10} {} 
    \NewMathAlphabet{mathbfss} {cmssbx10} {} 
  \fi
  \ifAMStwofonts
    \ifCUPmtlplainloaded \else
      \NewSymbolFont{upmath} {eurm10}
      \NewSymbolFont{AMSa} {msam10}
      \NewMathSymbol{\upi}     {0}{upmath}{19}
      \NewMathSymbol{\umu}     {0}{upmath}{16}
      \NewMathSymbol{\upartial}{0}{upmath}{40}
      \NewMathSymbol{\leqslant}{3}{AMSa}{36}
      \NewMathSymbol{\geqslant}{3}{AMSa}{3E}

      \let\leq=\leqslant 
      \let\geq=\geqslant \let\ge=\geqslant
    \fi
  \fi
\fi 

\ifnfssone
  \newmathalphabet{\mathit}
  \addtoversion{normal}{\mathit}{cmr}{m}{it}
  \addtoversion{bold}{\mathit}{cmr}{bx}{it}
  \newmathalphabet{\mathbfit} 
  \addtoversion{normal}{\mathbfit}{cmr}{bx}{it}
  \addtoversion{bold}{\mathbfit}{cmr}{bx}{it}
  \newmathalphabet{\mathbfss} 
  \addtoversion{normal}{\mathbfss}{cmss}{bx}{n}
  \addtoversion{bold}{\mathbfss}{cmss}{bx}{n}
  \ifAMStwofonts
    \ifCUPmtlplainloaded \else
      \UseAMStwoboldmath
      \makeatletter
      \new@mathgroup\upmath@group
      \define@mathgroup\mv@normal\upmath@group{eur}{m}{n}
      \define@mathgroup\mv@bold\upmath@group{eur}{b}{n}
      \edef\UPM{\hexnumber\upmath@group}
      \new@mathgroup\amsa@group
      \define@mathgroup\mv@normal\amsa@group{msa}{m}{n}
      \define@mathgroup\mv@bold\amsa@group{msa}{m}{n}
      \edef\AMSa{\hexnumber\amsa@group}
      \makeatother
      \mathchardef\upi="0\UPM19
      \mathchardef\umu="0\UPM16
      \mathchardef\upartial="0\UPM40
      \mathchardef\leqslant="3\AMSa36
      \mathchardef\geqslant="3\AMSa3E

      \let\leq=\leqslant 
      \let\geq=\geqslant \let\ge=\geqslant
    \fi
  \fi
\fi 

\ifnfsstwo
  \DeclareMathAlphabet{\mathbfit}{OT1}{cmr}{bx}{it}
  \SetMathAlphabet\mathbfit{bold}{OT1}{cmr}{bx}{it}
  \DeclareMathAlphabet{\mathbfss}{OT1}{cmss}{bx}{n}
  \SetMathAlphabet\mathbfss{bold}{OT1}{cmss}{bx}{n}
  \ifAMStwofonts
    \ifCUPmtlplainloaded \else
      \DeclareSymbolFont{UPM}{U}{eur}{m}{n}
      \SetSymbolFont{UPM}{bold}{U}{eur}{b}{n}
      \DeclareSymbolFont{AMSa}{U}{msa}{m}{n}
      \DeclareMathSymbol{\upi}{0}{UPM}{"19}
      \DeclareMathSymbol{\umu}{0}{UPM}{"16}
      \DeclareMathSymbol{\upartial}{0}{UPM}{"40}
      \DeclareMathSymbol{\leqslant}{3}{AMSa}{"36}
      \DeclareMathSymbol{\geqslant}{3}{AMSa}{"3E}

      \let\leq=\leqslant 
      \let\geq=\geqslant \let\ge=\geqslant
    \fi
  \fi
\fi 

\ifCUPmtlplainloaded \else
  \ifAMStwofonts \else 
    \def\upi{\pi}
    \def\umu{\mu}
    \def\upartial{\partial}
  \fi
\fi

\title
[The Ray Bundle Method]
{The Ray Bundle method for calculating weak magnification by
gravitational lenses}
\author
[C.J. Fluke et al.]
{C.J. Fluke, R.L. Webster \& Daniel J. Mortlock\\
School of Physics, The University of Melbourne, Parkville, Vic, 3052,
Australia}
\date{1998 December 16}
\pagerange{\pageref{firstpage}--\pageref{lastpage}}
\pubyear{1998}

\def\LaTeX{L\kern-.36em\raise.3ex\hbox{a}\kern-.15em
    T\kern-.1667em\lower.7ex\hbox{E}\kern-.125emX}

\begin{document}

\label{firstpage}

\maketitle

\begin{abstract}
We present here an alternative method for calculating magnifications in
gravitational lensing calculations -- the Ray Bundle method.  We provide
a detailed comparison between the distribution of magnifications obtained
compared with analytic results and conventional ray-shooting methods.
The Ray Bundle method provides high accuracy in the weak lensing limit, and 
is computationally much faster than (non-hierarchical) 
ray shooting methods to a comparable accuracy.

The Ray Bundle method is a powerful and efficient technique with which to
study gravitational lensing within realistic cosmological models, 
particularly in the weak lensing limit.
\end{abstract}


\section{Introduction}
Gravitational lensing is the study of the effects of matter on the propagation
of light.  The most obvious observational results are the production of
multiple images (as first seen with the multiply imaged quasar 0957+561
by Walsh, Carswell \& Weymann \shortcite{WALSH79}), the creation of
giant luminous arcs (first identified in the galaxy clusters Abell 370 
and Cl~2244 by Lynds \& Petrosian \shortcite{LYNDS86} and, independently,
in Abell 370 by Soucail et al. \shortcite{SOUCAIL87})
and the large magnifications of source flux seen in microlensing events
(for example, brightening of a single image of the multiply-imaged
quasar 2237+0305 due to compact objects in the lensing galaxy,
first detected by Irwin et al. \shortcite{IRWIN89}).

The paths of light rays from a source to the observer are conveniently  
described with the gravitational lens equation (\eqn{alenseqn} below).
This equation is highly non-linear, so that, 
except for a small number of specific cases,
there are no analytic solutions. In particular, 
there is no straightforward result which determines the 
image locations or magnifications for an ensemble of many lenses.  This 
presents a serious problem when we wish to study the lensing properties of
a complex lensing structure such as the Universe.  

Fortunately, a number of approximate numerical methods have been developed 
which  allow us to calculate magnifications and other properties of 
a collection of lenses.  Foremost amongst these are the Ray Shooting methods,
introduced by Paczy\'{n}ski \shortcite{PACZYNSKI86} and
Kayser, Refsdal \& Stabell \shortcite{KAYSER86} 
and developed by Schneider \& Weiss \shortcite{SCHNEIDER87},
Kayser et~al. \shortcite{KAYSER89} and Lewis et al. \shortcite{LEWIS93}.  
For a discussion of other methods, such as the use of
a scalar deflection potential, and the optical scalar equations, see 
Schneider, Ehlers \& Falco \shortcite{SCHNEIDER92}. 
 
In this paper, we present an alternative technique for calculating the 
magnification properties of an ensemble of lenses -- the Ray Bundle 
method (RBM).  The RBM is particularly well suited to studies 
of the weak lensing limit, where we are not concerned with the creation of
multiple (comparably bright) images.
Like the Ray Shooting method, the RBM uses backwards
propagation of light rays from the observer to the source, which are deflected
by the distribution of lenses, and are mapped to the source plane.  Whereas
the Ray Shooting method collects the deflected light rays within 
a rectangular grid of pixels,
we consider an infinitesimal bundle of rays which form a circular image, and 
maintain this association to the source plane.  

At first sight, the Ray Bundle method may seem to be less computationally 
efficient 
than the Ray Shooting method, as a grid based technique lends itself to
(fast) hierarchical calculations \cite{WAMBSGANSS90b}. However 
instead of requiring $\gesim 100$ light
rays per source grid to reduce the statistical error, we need only use
a bundle of $N_{\rm ray} = 8$ rays  to obtain magnifications which are 
correct to better than 5 per cent for an equivalent source size.  

With a Ray Shooting method, we have both image and source pixels, 
but cannot easily determine the correspondence between them.
By keeping track of the individual light bundles with the RBM, we are 
able to monitor the shape distortions of the beam caused by the 
shear and convergence of a lens ensemble.   This is of particular
interest when the RBM is applied to multiple lens plane geometries (as are
conventionally used for studying cosmologically distributed lenses).
Details of the beam shape allows for the opportunity 
to make comparisons between results calculated with the 
gravitational lens equation, and those using the optical 
scalar equations (which are more easily applied
to smooth mass distributions, or approximate mass distributions such as
the Swiss cheese model \cite{EINSTEIN45}).

In \sct{lensing} we introduce various basic results  of gravitational 
lensing, particularly with regards to magnification and the magnification
probability distribution.   The Ray Bundle method is introduced in
\sct{RayBundle} and compared in detail with analytic solutions for the
Schwarzschild lens model.   By obtaining the magnification probability 
distribution with both the Ray Bundle and Ray Shooting methods for a variety 
of lens geometries, we demonstrate the general
applicability of the RBM.  We do not discuss applications
of the RBM here, but reserve details for 
Fluke, Webster \& Mortlock (in preparation) where the Ray Bundle method is 
used to investigate weak lensing within realistic cosmological models 
(generated with N-body simulations).

\section{Gravitational Lensing}
\label{sct:lensing}

We present here a number of important results from gravitational lensing which 
we will require:  the gravitational lens equation, the magnification of
source flux and the magnification probability distribution.  
Several excellent sources exist which describe 
gravitational lensing
far more comprehensively than may be discussed here, for example
Schneider et~al. \shortcite{SCHNEIDER92} and 
Narayan \& Bartelmann \shortcite{NARAYAN96}.

\subsection{The gravitational lens equation}
\label{sct:lenseqn}
\begin{figure}
{\psfig{figure=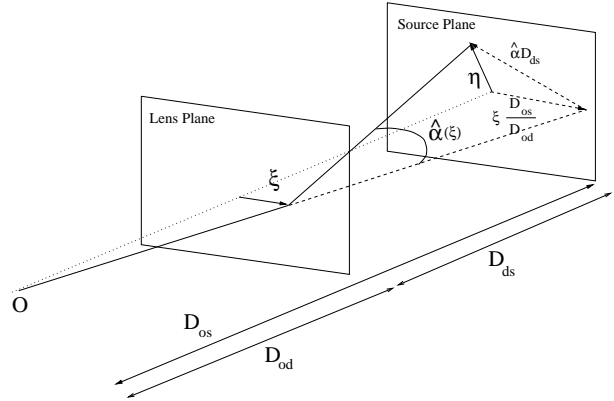,width=8cm}}
\caption{\label{fig:flenseqn}Geometrical arrangement for the 
gravitational lens equation for an observer at O.  $\bvec{\xi}$ is 
the impact parameter in
the lens plane, $\bvec{\eta}$ the sources position, and $\hat{\bvec{\alpha}}
(\bvec{\xi})$
the deflection angle of the light ray.   The distances ($D_{\rm ij}$) are
angular diameter distances between the [o]bserver, [d]eflector and [s]ource
planes.}
\end{figure}

The deflection of a light ray by a massive object is conveniently 
expressed with the gravitational lens equation (GLE), which may be derived 
from simple geometrical arguments (as demonstrated in \fig{flenseqn}).

The GLE relates the impact
parameter, $\bvec{\xi}$, in the lens (deflector) plane,
the source position, $\bvec{\eta}$, in the source plane and the deflection
angle, $\hatalpha(\bvec{\xi})$, of the light ray:
\begin{equation}
\bvec{\eta} = \frac{\Dos}{\Dod} \bvec{\xi} - \Dds \hatalpha
(\bvec{\xi}).
\label{eqn:alenseqn}
\end{equation}
The distances ($D_{\mbox{\small ij}}$) 
in \eqn{alenseqn} are angular diameter distances
between the [o]bserver, [d]eflector and [s]ource planes.  For a given
source position, an image will occur for each value of $\bvec{\xi}$ which 
is a solution of \eqn{alenseqn}.  We define the lens axis as the line from 
the observer through a single lens (at the origin of the lens plane), and which
is perpendicular to both the lens and source planes.

The GLE may be recast into a dimensionless form 
by introducing the scaling lengths $\xi_0$ and $\eta_0 = \xi_0 \Dos/\Dod$.
Defining $\bvec{x} = \bvec{\xi}/\xi_0$, $\bvec{y} = \bvec{\eta}/\eta_0$ and
the dimensionless deflection angle
\begin{equation}
\bvec{\alpha}(\bvec{x}) = \frac{\Dod\Dds}{\Dos \xi_0}\hatalpha 
(\xi_0 \bvec{x})
\end{equation}
it follows that
\begin{equation}
\bvec{y} = \bvec{x} - \bvec{\alpha}(\bvec{x}).
\label{eqn:Adlenseqn}
\end{equation}

\subsection{Magnification}
\label{sct:magstuff}
For a bundle of light rays passing through a transparent lens, 
the number of photons
is conserved, ie. gravitational lensing does not change the specific intensity
of the source.  A change in flux, however, can occur as the cross-sectional
area of a bundle of light rays will be affected by a gravitational lens.
The change in the apparent luminosity is entirely due to the change in the
solid angle that the image covers (at the expense of the rest of the sky) 
-- with a lens
present, more photon trajectories are brought to the observer's eye than
if there were no lens \cite{DYER81}.

If the flux at a frequency $\nu$ is $S_{\nu} = I_{\nu} d\Omega_{\rm obs}$,
where $I_{\nu}$ is the specific intensity and $d\Omega_{\rm obs}$ is the 
solid angle subtended by the source at the observer's location,
then the magnification is
\begin{equation}
\vert \mu \vert = \frac{S_{\nu}}{S_{\nu}'} = 
\frac{{\rm d}\Omega_{\rm obs}}{{\rm d}\Omega_{\rm obs}^{\prime}}
\label{eqn:solidAngle}
\end{equation}
with primes denoting quantities when the lens is absent.

It is usual to measure the magnification with respect to either a `full
beam' or an `empty beam'.  The full beam refers to the case where matter
is smoothly distributed everywhere (inside and outside of a bundle of 
light rays), so that there is no shear and the magnification is due 
entirely to convergence. 
An empty beam corresponds to the case where there is a smooth distribution
of matter (on average) external to the beam, and no matter within the beam. 
The empty beam is the maximally divergent beam, as there is now no
magnification due to convergence, and the
minimum magnification is $\mu_{\rm empty} = 1$. The addition of material 
outside of the beam will result in a total magnification $\mu_{\empty} \geq 1$
due to shear \cite{SCHNEIDER84}.  
In the work that follows, we will calculate magnifications
with respect to an empty beam.

\subsection{The magnification probability distribution}
\label{sct:MPH}
The (differential) magnification probability, $p(\mu,z) {\rm d}\mu$, 
is the probability that the total magnification of a source  at redshift
$z$ will lie in the range $\mu$ to $\mu + {\rm d}\mu$.  The probability is
subject to the constraints
\begin{equation}
\int_1^{\infty} p(\mu, z) {\rm d}\mu = 1
\label{eqn:normal}
\end{equation}
and
\begin{equation}
\int_1^{\infty} p(\mu, z) \mu {\rm d}\mu = \langle \mu \rangle(z),
\label{eqn:fluxcons}
\end{equation}
where $\langle \mu \rangle (z)$ is the mean magnification.  For a 
universe where the matter is smoothly distributed, 
$\langle \mu \rangle(z) = 1$, while for the empty beam magnifications
considered here, $\langle \mu \rangle(z) > 1$.  
\eqn{normal} provides the normalisation of the probability, while 
\eqn{fluxcons} expresses the conservation of flux \cite{WEINBERG76}.  

These constraints are only strictly true 
when we consider the magnification probability over the whole sky -- 
in the tests we conduct here, we use artificial lens distributions that
do not cover the entire celestial sphere.

The probability distribution provides a straightforward way of making a  
comparison between different techniques for solving the GLE.   Although we
might expect the calculation of the magnification along a particular 
line-of-sight to vary slightly (depending on source geometry, etc),
the statistical properties of a distribution of sources should be essentially 
independent of the method.

In practice, we will use a histogram of the probability as a function of
$\mu$.  This involves summing over a range of magnifications from
$\mu_1$ to $\mu_2 = \mu_1 + \Delta \mu$, so that each histogram bin represents
\begin{equation}
p(\mu_1,z) \Delta \mu = \int_{\mu_1}^{\mu_2} p(\mu,z)  {\rm d}\mu.
\label{eqn:mph_def}
\end{equation} 
In the following, we will refer to this as the magnification probability 
histogram (MPH). 

\subsection{Solving the gravitational lens equation}
The GLE is highly non-linear, and in general there are multiple 
solutions for the image locations for a single source position.  As a result,
analytic solutions (where we can invert the GLE to solve for all $\bvec{\xi}$ 
as a function of $\bvec{\eta}$) exist for only special lens geometries 
and models, which may not always be realistic or practical. 
Instead, various numerical methods have been developed which 
enable solutions to the lens equation to be found accurately and efficiently.

Some of the most widely used methods are those based on the 
concept of Ray Shooting. 
The basic principle of the Ray Shooting method  (RSM) is to use  \eqn{alenseqn} 
to propagate light rays 
backwards from the observer through  a sequence of one or more lens planes
to the source plane,
where the rays are collected on a rectangular pixel grid.  Typically $\sim 10^6$
pixels are required, with an average of $\bar{N} \sim 100$ rays 
per source plane pixel.  
The magnification in each source pixel $(i,j)$ is then proportional to 
the density of rays collected therein:
\begin{equation}
\mu_{\rm pixel}(i,j) = N_{\rm collected}(i,j)/\bar{N} .
\end{equation}

A shooting region in the lens plane is chosen such that only a few rays
from outside of this grid are mapped onto one of the sources, 
otherwise there would be missing flux.
By using a uniform grid of image rays, 
the relative error is approximately $\bar{N}^{-3/4}$ \cite{KAYSER86}, which
is better than the Poisson error if a random distribution of image rays was 
used (however a regular grid may introduce systematic errors).
As a consequence of this error, it is possible to 
get magnifications $\mu_{\rm pixel}(i,j) < 1$, violating the condition 
on the total magnification (\sct{magstuff}), 
which is entirely a numerical effect.   

Early versions of this method were introduced by Paczy\'{n}ski 
\shortcite{PACZYNSKI86} and 
Kayser et~al. \shortcite{KAYSER86} 
and developed by Schneider \& Weiss \shortcite{SCHNEIDER87}
and Kayser et~al. \shortcite{KAYSER89}.  
Hierarchical tree methods were applied to microlensing
scenarios by Wambsganss \shortcite{WAMBSGANSS90b} and
Wambsganss, Paczy\'{n}ski \& Katz \shortcite{WAMBSGANSS90a}. 
The hierarchical methods approximate the effects of lenses which are 
far from a light ray, and allow the inclusion of many thousands of lenses at a 
low computational cost (O($N\log_2 N$) for a tree-code with N lenses, 
versus O($N^2$) when the contribution of every lens is explicitly calculated).
An improvement on conventional Ray Shooting methods for obtaining 
statistical properties of microlensing light curves can be made with the
efficient one-dimensional contour following algorithm of Lewis et al. 
\shortcite{LEWIS93}.
We now introduce the Ray Bundle method as an alternative to the RSM.

\section{The Ray Bundle Method}
\label{sct:RayBundle}
The Ray Bundle method (RBM) is similar to the RSM, in that the lens
equation is used to propagate light rays backwards from the observer to the
source plane.  However we now consider a bundle consisting of a central
ray (the null geodesic) surrounded by $N_{\rm ray}$ light rays,
which create an image shape (usually circular).   As the ray bundle
passes through the lens distribution, its shape will be distorted due to shear
(stretching along an axis) and convergence (focusing due to matter 
within the beam).  
For an `infinitesimal' ray bundle, the magnification is determined
from \eqn{solidAngle} by calculating the area of the bundle in the image plane
($d \Omega_{\rm obs}^{\prime}$) and the source plane ($d \Omega_{\rm obs}$).

Since we are using backwards ray-tracing through a single image position, 
we do not know where other images may occur -- hence any measurement of the 
magnification using the RBM will underestimate the total magnification. 
However, this is only a significant problem when we consider images
located near the critical curves, when the contribution to the total 
magnification due to any other images becomes important (see below). 
The RBM was developed for applications in the weak lensing limit, and 
should be used with caution for strong lensing cases.

\subsection{Comparison with analytic solutions}
\label{sct:comparison}
We can investigate the validity of the RBM by comparison with the
various analytic solutions which exist for the Schwarzschild lens (see
\app{SLens} for a summary).

Consider first a circular source of radius $R_{\rm s}$ with 
centre at $\bvec{y}_{\rm c} = (y_{1, \rm c}, y_{2, \rm c})$.  
The circumference of the source is then described by the set of vectors
$\bvec{y} = (y_1, y_2)$ with
\begin{eqnarray}
y_1 &=&y_{1, \rm c} + R_{\rm s} \cos (\phi) \nonumber \\
y_2 &=& y_{2, \rm c} + R_{\rm s} \sin (\phi) 
\end{eqnarray}
where $0 \leq \phi < 2 \pi$.
 
For each $\bvec{y}$, we can solve for the two solutions, $\bvec{x}_{\pm}$,
with \eqn{xpm}. In this case we are using the GLE to map from the source
plane to the image plane.  A source far from the lens axis produces one highly 
demagnified image ($\mu_{\rm faint}$) located near the lens axis (at
$\bvec{x}_{\rm faint}$).  The second
image will have a magnification $\mu_{\rm bright} \geq 1$, and an angular
position near the source (at $\bvec{x}_{\rm bright}$).
As the source is moved towards the lens axis, the images are 
stretched in the tangential direction\footnote{The Schwarzschild lens 
has a single (degenerate) caustic point at $y = 0$, which
corresponds to a tangential critical curve at $x = 1$ (the Einstein ring).}
and become comparable in brightness ($\vert \mu_{\rm faint}\vert 
\sim \vert \mu_{\rm bright} \vert \gg 1$
as $\bvec{y}_c \rightarrow 0$).  When $\bvec{y}_c = 0$, the two images 
merge into a highly magnified ring (the Einstein ring) 
with total magnification given by \eqn{muemax}.

\begin{figure}
\hspace{1.3cm}{\psfig{figure=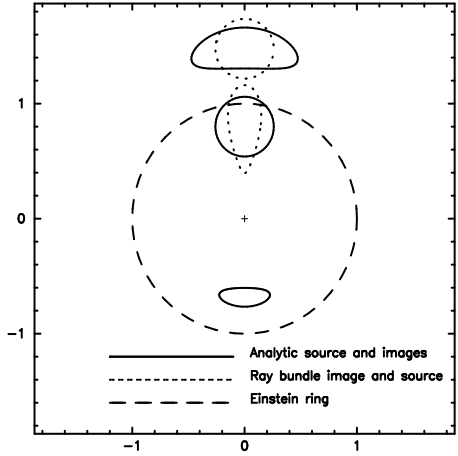,height=6cm,width=6cm}}
\caption{\label{fig:impos} Shapes and locations 
of a circular source, for which the boundary points may be determined for
both images (solid line), and a circular image and corresponding source
(short dashed line).  The lens is located at the cross, and the long dashed
line is the Einstein radius.  In both cases (solid and short dashed line) 
the source is the shape which intersects the Einstein radius. 
The scale is in units of the Einstein radius.}
\end{figure}

Now consider a circular image of radius $R_{\rm i}$ centred on the location
of the bright image, 
$\bvec{x}_{\rm c} = \bvec{x}_{\rm bright} (\bvec{y}_{\rm c})$, 
with circumferential points $\bvec{x} = (x_1, x_2)$
\begin{eqnarray}
x_1 &=&x_{1, \rm c} + R_{\rm i} \cos (\phi) \nonumber \\
x_2 &=&x_{2, \rm c} + R_{\rm i} \sin (\phi) .
\end{eqnarray}
This time, the GLE maps in the opposite direction -- from the image
plane to the source plane.  The source shape we obtain is stretched along the 
radial direction, and differentially compressed in the tangential direction.

\fig{impos} demonstrates the differences between the shape and locations 
of a circular source (solid line), for which the circumferential points may be 
determined for both images,
and a circular image and its corresponding source
(short dashed line).  In this example, we have considered a source which is
near the Einstein radius where strong lensing effects dominate, and there
may be a significant contribution to the flux from the second image.
In all cases where we will apply the RBM, the image is chosen to be 
well away from the Einstein radius (critical curve), and so 
the flux lost from the second image is not important, as we now show.  

\subsection{The magnification deficit}
\label{sct:deficit}
We now look at how accurately the RBM approximates the total magnification,
even though it includes the contribution of only one image.  
If we set $R_{\rm s} = R_{\rm i}$, as we expect only small changes to the shape and hence radius of the image in the weak lensing 
limit ($\vert \bvec{x}_{\rm c} \vert \gg 1$), then this is a two 
parameter problem ($R_{\rm i}, N_{\rm ray}$).

Defining the RBM magnification in terms of the 
ray bundle image and source areas ($A_{\rm i, RBM}, A_{\rm s, RBM}$) as
\begin{equation}
\mu_{\rm RBM} = \frac{A_{\rm i, RBM}}{A_{\rm s, RBM}}  
\end{equation}
and the true (total) magnification as
\begin{equation}
\mu_{\rm true} = \frac{A_{\rm faint} + A_{\rm bright}}{A_{\rm s}},
\label{eqn:areatot}
\end{equation}
where $A_{\rm faint}$, $A_{\rm bright}$ are the areas of the two images,
then the relative error in $\mu_{\rm RBM}$ is
\begin{equation}
\frac{\Delta \mu_{\rm RBM}}{\mu_{\rm true}}
= \frac{\vert \mu_{\rm true} - \mu_{\rm RBM} \vert}
{\mu_{\rm true}}.
\end{equation}

Due to the circular symmetry of the Schwarzschild lens model, we need only
determine the radius, $x_{\rm cut}$, within which the RBM produces a
relative error $\frac{\Delta \mu_{\rm RBM}}{\mu_{\rm RBM}} > p$ per cent.

By using $N_{\rm ray}$ rays in the image and source bundles, we are 
approximating the shape of a circular image/source by a polygon with 
$N_{\rm ray}$ sides.  Clearly, when $N_{\rm ray} \gg 1$, we will have 
a reasonable approximation to the true shape of the image/source. However,
to improve the speed of the ray bundle method (at the cost of a small error),  
we ideally want to select a small value of $N_{\rm ray}$ ($\lesim 20$). 
The areas are
calculated as a sum of triangular components within the image/source
polygon, where each triangle  has a common vertex at $\bvec{y}_{\rm c}$ or 
$\bvec{x}_{\rm c}$ (ie. the null geodesic).  

We need to first check that the calculated $\mu_{\rm true}$ is not 
significantly in error using a particular value of $N_{\rm ray}$.  This was 
achieved by numerically integrating \eqn{analytic}, and comparing with
$\mu_{\rm true}$ for $4 \leq N_{\rm ray} \leq 256$.  
The relative error in $\mu_{\rm true}$ was found to be independent of
the choice of $N_{\rm ray}$,  and was well below the percentage cut-off 
level selected for $\mu_{\rm RBM}$ at the same radii.  

\begin{figure}
\hspace{1.1cm}{\psfig{figure=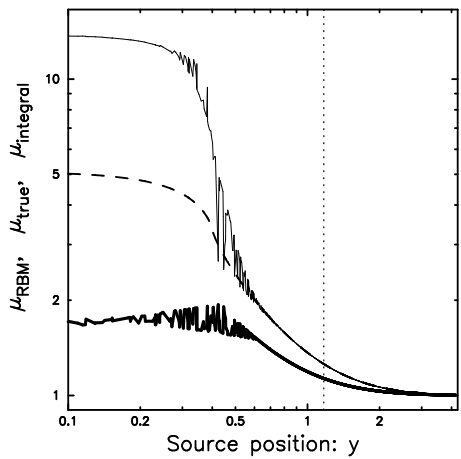,height=6cm,width=6cm}}
\caption{\label{fig:xcut} Magnification of an extended source 
($R_{\rm s} = 0.4$), near a Schwarzschild lens.  
$\mu_{\rm RBM}$ (thick solid line) is the
Ray Bundle method magnification based on a single image; $\mu_{\rm true}$
(thin solid line) is the total magnification for both images and 
$\mu_{\rm integral}$ (dashed  line) is the solution of \eqn{analytic}.
$\mu_{\rm RBM}$ is accurate to $10$ per cent at $y_{\rm cut} = 1.17$, 
as indicated by the vertical dotted line. }
\end{figure}

For source positions which are near the
lens axis, it is difficult to  keep track of which solutions $x_{\pm}$ 
belong to which of the images, particularly when merging of images is
taking place.  In these cases, $\mu_{\rm true}$ is a misnomer, as it
can be significantly higher than the total magnification from \eqn{analytic},
as may be seen in \fig{xcut}.
$\mu_{\rm RBM}$ is subject
to a similar error, as the areas here are also being calculated on the basis
of $N_{\rm ray}$ rays.  We can proceed under the assumption 
that $\mu_{\rm true}$ is sufficiently accurate for comparison with 
$\mu_{\rm RBM}$ when the two images are separable.

By randomly selecting $5000$ source (RBM image) locations, we calculate
$\mu_{\rm true}$ and $\mu_{\rm RBM}$.  \tbl{fluxlost} shows the 
dimensionless radius, $x_{\rm cut}$, within
which $\Delta \mu_{\rm RBM}/\mu_{\rm true} > p$~per~cent for a range of 
source radii ($10^{-5} < R_{\rm i} < 0.01$).  
For a given value of the relative error, these results are essentially 
independent of both the source radii and $N_{\rm ray}$ in the
range $4 \leq N_{\rm ray} \leq 256$.  

The choice of $N_{\rm ray} = 8$ appears to be the best compromise between 
accuracy and speed for our later investigations of the RBM.  The smaller $N_{\rm ray}$, 
the fewer total deflection calculations which need to be made, yet we
retain high accuracy for $\mu_{\rm RBM}$ in the weak lensing limit.   
With 8 rays, we have a symmetric image with a quadrupole component, so 
it is possible to determine the distribution of image ellipticities.

In \fig{xcut}, we plot $\mu_{\rm RBM}$ (thick solid line), 
$\mu_{\rm true}$ (thin solid line) and $\mu_{\rm integral}$, the solution
of \eqn{analytic} (dashed line), as functions of the source impact parameter 
($y = \vert \bvec{y} \vert$).  
The source/image bundle radius in this case was $R_{\rm s} = R_{\rm i} = 
0.4$.  We have used a larger radius than would be practical for the RBM in 
order to show in more detail what the high magnification
behaviour is like near $y=0$.  As the source is comparable in size
to the Einstein radius, the calculated magnifications near $y \simeq 0.4$ 
are noisy.  In this case, a circular source should produce a pair of 
highly distorted arcs, but this is not well represented by the use of triangular
components within the ray bundle (ie. parity changes can occur in the individual
triangles which make up the bundle). 
The vertical dotted line shows where the relative error in $\mu_{\rm RBM}$
is accurate to $10$ per cent, at $y_{\rm cut} = 1.17$ or $x_{\rm cut} = 1.74$ 
(note that these values are higher than in \tbl{fluxlost} due to the larger
source radius).

\begin{table}
\caption{\label{tbl:fluxlost} Position of source, $y_{\rm cut}$, at which 
the relative error in $\mu_{\rm RBM}$ (compared to the true magnification) 
is first $> p$~per~cent, caused by neglecting flux from second image.   
$x_{\rm cut}$ is the corresponding (bright) image position. Values in the table
are for source radii $10^{-5} < R_i < 0.01$ and $4 \leq N_{\rm ray} \leq 256$.} 
\begin{center}
\begin{tabular}{lrrrrr}
& \multicolumn{5}{c}{$\Delta \mu_{\rm RBM}/\mu_{\rm true} = p\%$}\\
 & $1\%$ & $2\%$ & $5\%$ & $10\%$ & $20\%$ \\ 
\hline
$y_{\rm cut}$ & 2.86 & 2.26 & 1.61 & 1.15 & 0.70\\
$x_{\rm cut}$  & 3.17 & 2.64 & 2.09 & 1.73 & 1.41  
\end{tabular}
\end{center}
\end{table}

\begin{figure}
\begin{center}
{\psfig{figure=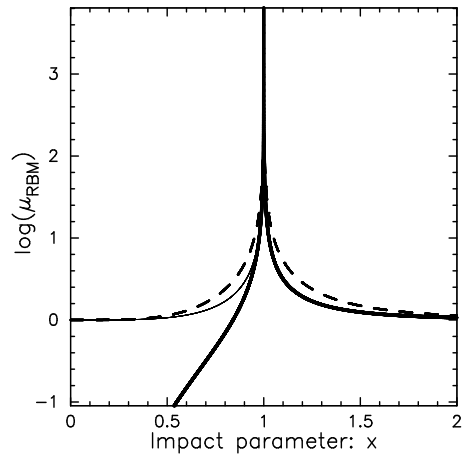,height=6cm,width=6cm}}
\end{center}
\caption{\label{fig:bundlemu} Magnification as a function of
the impact parameter to a Schwarzschild lens.  The thick solid line
is the magnification for the RBM, the dashed line is for a point source
with one image at the same location as the RBM image, and the thin solid
line is the `corrected' RBM magnification: $\mu^{\prime} = \mu_{\rm RBM} +1$.
The Einstein radius corresponds to $x = 1$, the image radius is $R_{\rm i} =
10^{-4}$ and $N_{\rm ray} = 8$.}
\end{figure}

This test has shown that the Ray Bundle method can give a highly accurate 
value of the magnification (relative error better than $5$ per cent) in 
the weak 
lensing limit.  However, we have modified the Ray Bundle method slightly by
requiring that the image position agrees with the brighter image 
of the corresponding circular source.  In a true application of the RBM, we 
randomly select image positions without the {\em a priori} knowledge of 
source locations.  

\fig{bundlemu} shows the RBM magnification for bundle positions near,  and
within, the critical curve of a single Schwarzschild lens (thick solid line).  
For comparison, the total magnification of a point source with one image
at the same location as the RBM image is shown as the dashed line.  
Bundle positions within $x=0.5$ produce magnifications 
$\mu_{\rm RBM} \ll 1$.  For these bundles we actually selected
the fainter of the images, as opposed to the case considered 
previously, where we purposefully selected the bundle corresponding to the 
brighter image.  Solutions of the lens equation, \eqn{xpm}, 
which produce an image near
the lens ($x_- \lesim 1$) correspond to total magnifications near $\mu = 1$, 
as  the source is far from the caustic point ($y \simeq - 1/x_- > 1$).  
This is demonstrated with the thin solid line in \fig{bundlemu}, where
we plot ray bundles with $\mu \leq 1$ 
within the Einstein radius as $\mu^{\prime} = \mu + 1$.  

It is clear that a small strip of image locations near the critical curve
is responsible for the high magnification region of the MPH, for which the
Ray Bundle method is not well suited.  

Images well within the Einstein
radius correspond to sources far from the lens axis.  Such images are the
faint images described in \sct{comparison} (at $\bvec{x}_{\rm faint}$), 
and so there will be a second image
at $\bvec{x}_{\rm bright}$ which contributes the majority of the flux.  
Although it is not possible to solve for $\bvec{x}_{\rm bright}$ given 
$\bvec{x}_{\rm faint}$, if the entire image plane is well sampled with 
ray bundles, then we can expect that another bundle will pass 
through $\bvec{x}_{\rm bright}$ and the source magnification will be
calculated on the basis of this second bundle only.

Two restrictions are now imposed on the Ray Bundle
method for its later application to ensembles of lenses, 
which serve to complete the definition of the method.  
Firstly, image positions 
within the Einstein radius, or equivalently, any magnifications which are
calculated to be $\mu < 1$ are discarded.  Secondly, after selecting a 
relative error
for $\mu_{\rm RBM}$, we do not include images which fall within $x_{\rm cut}$
Einstein radii of any given lens.

Having shown that in the `weak' lensing limit ($\vert \bvec{x}_{\rm c} 
\vert \gesim 2$), we can be 
sure of calculating the magnification to within $5$ per cent (or better) using
$N_{\rm ray} \ge 4$ we can now proceed to a statistical comparison between
the Ray Shooting and Ray Bundle methods.  

\subsection{Comparison of magnification probability histograms}
\label{sct:Numerical}
Next we compare the MPH obtained with the RBM and the RSM.  
A number of subtle differences exist between the two methods, even when 
applied to the same lens model. 
The source size investigated is limited by the number of pixels in the source
plane for the RSM.  For a grid of $N_{\rm pix} \times N_{\rm pix}$ 
covering a square region $2 y_{\rm max} \times 2 y_{\rm max}$, the source
`radius' is $R_{\rm s} = y_{\rm max}/N_{\rm pix}$.  
 
We choose the RSM sources to be squares with a side-length equal to the 
diameter of the circular RBM image bundles.  Since
the magnification decreases as the source area is increased, we expect
each RSM source to have a systematically smaller magnification than the
corresponding RBM image.
It is possible to use a much finer resolution grid for the ray shooting, and
then integrate over a larger source size, but we have elected not to do this.
This decision was based on a comparison of the computation time:  for 
our implementation of the RSM, 
a grid of $1000 \times 1000$ pixels, with an average of $\bar{N} = 250$ rays 
per pixels took approximately eight hours of computation 
time.  An equivalent number of images (where $N_{\rm ray} =8$) 
is completed in 1 minute with the RBM\footnote{More computationally 
efficient versions of the RSM are available, which can produce the 
same level of resolution in a time comparable to that of the RBM 
(J. Wambsganss, private communication). }

\begin{figure}
\hspace{1.3cm}{\psfig{figure=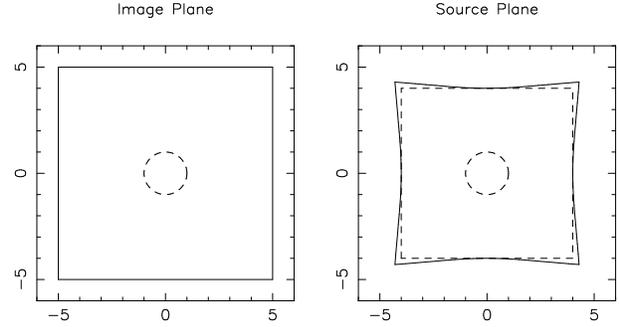,width=8cm}}
\caption{\label{fig:distort} A rectangular boundary in the image plane is
distorted in the source plane.  For the RSM, we only select sources in a 
regular grid which lies completely inside the distorted boundary, as shown by
the dashed square in the right-hand panel. 
The Einstein radius is shown for scale. }
\end{figure}

Due to the distortion of rays near the boundary, as shown in \fig{distort},
we actually shoot rays with the RSM through a larger angular region in the
image plane.  This prevents us from including source pixels which are not
well sampled by rays.

The accuracy of the RSM magnifications depends on the average number of
rays collected in each pixel, $\bar{N}$.  Since we are not implementing
a (fast) hierarchical  method, the time required to obtain the magnification
distribution is proportional to the total number of rays,
$N_{\rm RSM} = \bar{N} N_{\rm pix}^2$.
The RSM does not fully sample the highest magnification region 
$(\mu \approx \mu_{\rm max})$ due to the regular placement of sources on a 
grid. 

With the RBM image size fixed by $R_{\rm i} = R_{\rm s}$, 
we are left to 
choose the number of rays which make up the ray bundle, $N_{\rm ray}$, and
the number of images.  Ideally we want the images to completely cover the
image plane (which again has a slightly larger angular size than the source
plane due to the deflection of light rays near the boundaries) which requires
\begin{equation}
N_{\rm image} \approx f \frac{\pi R_{\rm i}^2}{(2 y_{\rm max})^2},
\end{equation}
where $f$ is (approximately) the fraction of the source plane covered 
by sources.
For the RBM then, the total number of rays required is 
$N_{\rm RBM} = N_{\rm image} \times N_{\rm ray}$.

The comparative computational speed and resulting magnification accuracy 
may be obtained by requiring $N_{\rm RBM} = N_{\rm RSM}$.  Setting $f = 1$
and $R_{\rm i} = R_{\rm s}$, we have 
\begin{equation}
\bar{N} = \frac{4}{\pi} N_{\rm ray}.
\end{equation}
We have already seen that $N_{\rm ray} = 8$ is a suitable choice, which gives
$\bar{N} \sim 10.2$.   This corresponds to a relative error $\bar{N}^{-3/4}
\sim 17$ per cent. 
Although we can relax the constraint on the fraction of the source plane
covered somewhat, and still have a well sampled MPH with the Ray Bundle 
method, we do not have this flexibility with the grid based Ray Shooting 
method.  In addition, as we decrease the source size, the number of pixels
required for the RSM increases, and a higher density of rays is necessary.  

\begin{figure}
\hspace{1.3cm}{\psfig{figure=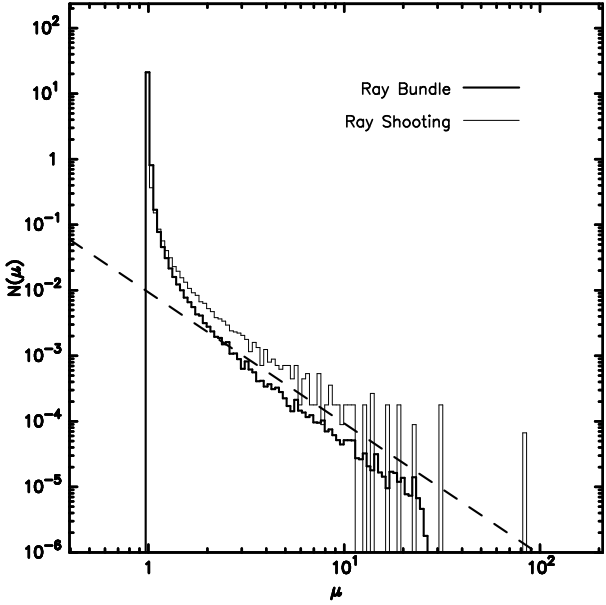,height=8cm,width=8cm}}
\caption{\label{fig:mucomp} Comparative magnification probability 
histograms for 
the Ray Bundle method (thick solid line) and the Ray Shooting method 
(thin solid line) for
a single Schwarzschild lens.  The dashed line shows the expected 
high magnification power law slope $N(\mu) \propto \mu^{-2}$.
See \tbl{comp_param} for the model parameters.}
\end{figure}

\fig{mucomp} shows the  Magnification Probability
histograms obtained for a single Schwarzschild lens using the Ray Shooting
(thin line) and Ray Bundle (thick line) methods.  The parameters for each
method are listed in \tbl{comp_param}.   The vertical axis of this (and later
histograms) is the normalised number of bundles/sources in each magnification
bin, $N(\mu)$, which is equivalent to the definition of $p(\mu_1)\Delta\mu$ in 
\eqn{mph_def}. 

\begin{table}
\caption{\label{tbl:comp_param} Parameters for Magnification probability
histogram comparison between Ray Bundle and Ray Shooting methods.  
$\mu_{\rm min}$ and $\mu_{\rm max}$ are the minimum and maximum magnifications
calculated for each method, $N_{\rm(b,s)}$ is the number of bundles/sources
used for the RBM/RSM histogram, $R_{\rm(b, s)}$ is the radius of a  
circular bundle for the RBM and the half pixel size for the RSM.
Results are given for a single Schwarzschild lens (see \fig{mucomp}) and 
a distribution of five Schwarzschild lenses (see \fig{mucomp5}).
In both case, $\bar{N} = 250$ rays for the RSM and $N_{\rm ray} = 8$ 
for the RBM. }

\begin{center}
\begin{tabular}{ccrrrc}
Method & $N_{\rm lens}$ & $(\mu_{\rm min}-1)$ & $\mu_{\rm max}$ & $N_{\rm(b,s)}$ & Radius
\\ \hline
RSM & 1 & --0.03     & 86.0 & $10^6$ & 0.01 \\
RBM & 1 & $2.4\!\times\!10^{-5}$ & 25.6 & $9.8\!\times\!10^5$& 
0.01 \\
RSM & 5 &  --0.094     & 37.7 & $10^6$ & 0.01 \\
RBM & 5 & 0.0  & 218.1 & $9.3\!\times\!10^5$& 0.01
\end{tabular}
\end{center}
\end{table}

A cut-off was imposed on
image locations for the RBM at $x_{\rm cut}=1.01$ Einstein radii.  The two 
distributions are qualitatively very similar, when the various caveats
described above are considered.  
Imposing a larger value of $x_{\rm cut}$ serves to 
reduce the maximum magnification with the RBM.  The poor sampling in
the highest $\mu$ bins for the RSM is clearly demonstrated.   The dashed
line shows the expected $\mu^{-2}$ power law slope of the MPH at large $\mu$
(see \app{SLens}), and both distributions have this approximate form (although
for the RSM the statistical significance of the histogram bins with $\mu >10$
is low).  

As expected, the RSM produces magnifications which are $\mu < 1$ (when 
$\mu_{\rm empty} \geq 1$ is expected) due to numerical effects.
The sample mean and variance of the two distributions are 
$\langle\mu\rangle = 0.98$ and $\sigma^2_{\mu} = 0.03$ for Ray Shooting, and
$\langle\mu\rangle = 1.02$ and $\sigma^2_{\mu} = 0.08$ for the 
Ray Bundle method\footnote{We reiterate that in the definition of the RBM, we
do not include images with $\mu < 1$, see end of \sct{deficit}). }.
The Ray Shooting method provides higher accuracy at 
high magnifications, but at magnifications $\mu \sim 1$, the Ray Bundle method
is more accurate {\em even though} flux from additional 
images is neglected. 

\begin{figure}
\hspace{1.3cm}{\psfig{figure=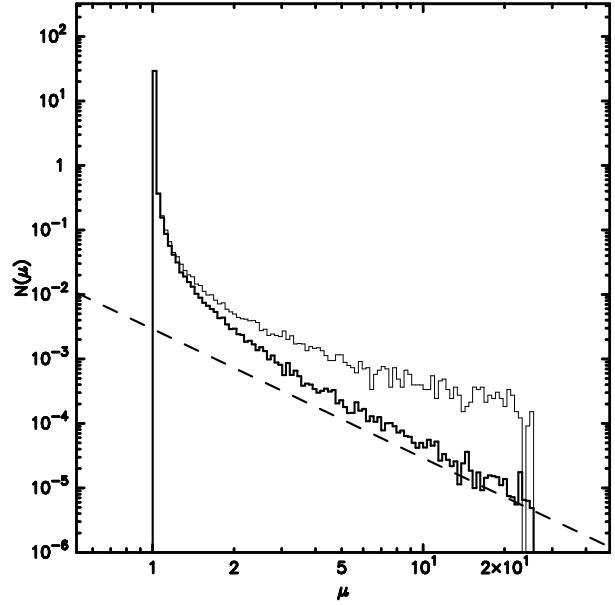,height=8cm,width=8cm}}
\caption{\label{fig:weight} Dependence of the Ray Bundle method magnification
probability histogram on the weighting.  
With no weighting (thin solid line) the MPH is incorrect at high 
magnification, where
the source sizes are much smaller than in the weak lensing limit, while
weighting by source area (thick solid line) gives the expected power law
slope \(N(\mu) \propto \mu^{-2}\) at $\mu > 10$ (dashed line). 
See \tbl{comp_param} for the RBM model parameters.}
\end{figure}

One aspect of the MPH we have not yet discussed has to do with the weighting
we apply to each ray bundle.  For the RSM, every source is a pixel with the
same area.  For the RBM, the initial ray bundles (images) have the same
area, but the resulting sources must have different areas (by the definition of
a magnification).  It is sufficient to weight each ray bundle in the MPH by 
the area of the resulting source.  \fig{weight} shows the 
RBM (thin line) with no weighting compared with the correct area weighting
(thick line).

\subsection{More complex lens distributions}
\label{sct:complex}
For an ensemble of $N$ lenses in the lens plane, each with mass $M_j$,
the total deflection angle generalises to
\begin{equation}
\hatalpha(\bvec{\xi}) = \sum^N_{j=1} \frac{4 G M_j}{c^2}
\frac{(\bvec{\xi} - \bvec{\xi}_j)}{\,\,\vert \bvec{\xi} - \bvec{\xi}_j \vert^2},
\end{equation}
where the $(\bvec{\xi} - \bvec{\xi}_j)$ are the impact parameters to each lens.

Consider the case where lenses are restricted to lie within a rectangular 
region\footnote{This is the case in common implementations of Ray Shooting
as it allows for the easy implementation of fast, hierarchical 
methods \cite{WAMBSGANSS90b}.}.
Light rays passing through one of the corners of the shooting region
will necessarily be deflected inwards by the mass distribution.  This is
appropriate for an isolated configuration of lenses, such as in studies of the 
microlensing effect of many stars which make up a galaxy (where the 
contribution to the deflection by an external mass distribution may be 
modelled by adding a shear term to the lens equation). However, for an
investigation of the lensing due to large scale structure, where the mass
distribution is assumed to be continuous and homogeneous in all directions
about a ray, we may introduce an artificial shear on rays near the 
shooting boundary.

For the RBM, we choose to calculate each of the deflection angles 
explicitly, with an increase in the computational time over an 
equivalent hierarchical method.  By making the direct calculation 
of the deflection,  we are free to choose the geometry of the region 
within which 
we include lenses.  The most natural choice for a distribution which is
homogeneous and isotropic beyond some length scale $R_{\rm H}$ is 
to include lenses within a circular region around each ray
out to the radius $R_{\rm lens} = R_{\rm H}$.    
For the isolated lens geometries we examine here, we need only set 
$R_{\rm lens}$ to encompass the RSM image plane.  A discussion of the 
appropriate choice for $R_{\rm H}$ in cosmological lensing scenarios is
reserved for Fluke et al. (in preparation).

\begin{figure}
\hspace{1.3cm}{\psfig{figure=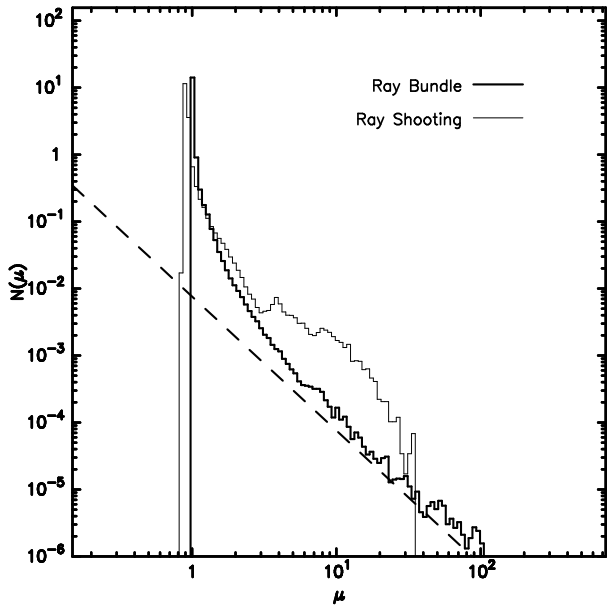,height=8cm,width=8cm}}
\caption{\label{fig:mucomp5} Comparative magnification histograms for 
the Ray Bundle method (thick solid line) and the Ray Shooting method 
(thin solid line) for a distribution of 5 equal-mass Schwarzschild lenses.  
The dashed line shows the expected high magnification power law slope 
$N(\mu) \propto \mu^{-2}$.  See \tbl{comp_param} for the model parameters.}
\end{figure}

As a final test here, we now consider a random distribution of 
$N_{\rm lens} = 5$ lenses (with the same lens positions for both the RSM
and RBM).  The resulting MPH is shown in \fig{mucomp5} and various parameters
in \tbl{comp_param}.
For both methods we use a total of $10^6$ bundles/sources, 
but discarding bundles within $x_{\rm cut} = 1.01$ Einstein radii 
for the RBM reduces 
this to $9.3\times10^5$ bundles for the histogram.   
The sample mean and variance are $\langle\mu\rangle = 1.09$, 
$\sigma^2_{\mu} = 2.09$ for the RBM and
$\langle\mu\rangle = 0.98$, $\sigma^2_{\mu} = 0.23$ for the RSM.

Using an ensemble of lenses introduces a new length scale to the problem,
so that sub-structure at intermediate magnifications ($3 \simeq \mu \simeq 30$)
is seen in the MPH using the RSM (but not with the RBM).  
The `bump' in the MPH occurs for a planar distribution of lenses, 
and has been studied both numerically \cite{RAUCH92} and analytically
\cite{KOFMAN97}, and is believed to be due to the caustic patterns of pairs 
of point lenses.  If this is a caustic-induced feature, ie. a region of 
high magnification, then the RBM will not provide the `correct' magnification. 
We feel the feature may be in part due to the low resolution with which 
the complex caustic structure was mapped with the RSM 
on a regular grid ($1000 \times 1000$ pixels), however a discussion of this
effect is beyond the scope of this paper.

\section{Conclusions}
\label{sct:Disc}
The Ray Bundle method provides a computationally fast, accurate and
flexible alternative to the Ray Shooting method for studies of the 
weak gravitational lensing limit.  A wide variety of lens models are
easily incorporated --  we have considered here only the case of 
the Schwarzschild lens, but changing to a different model involves 
modifying only $\hat{\bvec{\alpha}}(\bvec{\xi})$ in the GLE.
One alternative to ray based methods requires solving for the
scalar lensing potential, but it may not always be possible to find an
analytic solution for all lens models. 
The RBM also allows us to avoid artificial shear introduced by grid
based methods for light rays which pass near the shooting boundary, when
a small portion of an otherwise homogeneous and isotropic distribution
is used.  

The point source limit may be approached as values of $R_{\rm i}$ may be
selected without the restriction of introducing a finer source grid,  and
a corresponding increase in the density of rays required (provided we can relax
the constraint that the source plane must be completely covered by sources
with the RBM).

The RBM should only be applied with caution to strong lensing
scenarios, where the RSM is far superior.  As only one image
is followed to the source, there will be an error in the total magnification
when the image is near a critical curve. 

The RBM is, however,  particularly well suited to problems were we 
want to investigate in detail individual lines-of-sight for various 
lens geometries and models. An important advantage
of the RBM is that we can associate a particular image position and shape with 
the corresponding source position and shape.  This provides us with the 
opportunity of following the development of the shape of a ray bundle 
through a sequence of lens planes, as used in models of cosmological lensing
(for example, Wambsganss, Cen \& Ostriker \shortcite{WAMBSGANSS98}).  

An important application of weak lensing is to determine the effect of
small changes in the magnification of standard candle sources (such as Type
Ia Supernovae) on the derived values of cosmological 
parameters \cite{WAMBSGANSS97}.
The high accuracy of the RBM in the weak lensing limit makes 
it a value tool for such studies.

\section*{Acknowledgments}
The authors would like to thank Peter Thomas and Andrew Barber  
(University of Sussex), and Hugh Couchman (University of Western Ontario)
for helpful discussions.  The authors are grateful to the referee, Joachim
Wambsganss, for his insightful comments.
CJF and DJM are funded by Australian Postgraduate
Awards.  CJF is grateful for financial assistance from the 
University of Melbourne's Melbourne Abroad scholarships scheme, 
and the Astronomical Society of Australia's travel grant scheme.

\appendix
\section{The Schwarzschild Lens}
\label{app:SLens}
The simplest lens model is the point-mass or Schwarzschild lens 
(for example Schneider et al. \shortcite{SCHNEIDER92}, Narayan \&
Bartelman \shortcite{NARAYAN96}) , for
which the deflection angle due to a mass, $M$, is
\begin{equation}
\hatalpha(\bvec{\xi}) = \frac{4 G M}{c^2 \vert \bvec{\xi} \vert^2} \,
\bvec{\xi}.
\end{equation}
The dimensionless lens equation for a point source with this model is 
\begin{equation}
y = x - 1/x
\end{equation}
so that there are two images (one located on either  side of the lens,
and co-linear with the lens) at
\begin{equation}
x_{\pm} = \frac{1}{2}\left( y \pm \sqrt{y^2 + 4}\right)
\label{eqn:xpm}
\end{equation}
with corresponding magnifications
\begin{equation}
\mu_{\pm} = \frac{1}{2} \left(\frac{y^2 + 2}{y \sqrt{y^2 + 4}} \pm 1 \right).
\end{equation}
The total magnification is the sum of the absolute values
of the individual image magnifications: $\mu_{\rm p} = \vert \mu_+ \vert
+ \vert \mu_- \vert$.
When the source is far from the lens axis ($y \gg 1$), one of the images
(say, $x_-$) will be significantly demagnified ($\mu_- \ll 1$).
The total magnification is then $\mu_{\rm p} \approx \mu_{\rm +}$.

It is possible to derive an analytic form for the magnification probability,
$p(\mu, z)$, for the Schwarzschild lens for large
values of $\mu_{\rm p}$ \cite{SCHNEIDER92}, which is reasonably generic
for most lens models \cite{PEACOCK82}:
\begin{equation}
p(\mu) \propto \mu_{\rm p}^{-3}.
\label{eqn:mu3}
\end{equation}
On integrating to form the MPH we have
\begin{equation}
p(\mu) \Delta \mu = \int_{\mu_1}^{\mu_2} p(\mu) {\rm d}\mu =
\frac{1}{\mu_1^2} \left(1 - \frac{\mu_1^2}{\mu_2^2}\right).
\end{equation}
If the histogram bins are equally spaced logarithmically,
$\mu_1/\mu_2$ is constant, and so we have an additional constraint that
the MPH must have a power law slope $-2$ for $\mu \gesim 10$.

For an extended source, the
total magnification is the integral of $\mu_{\rm p}$ over the
source, weighted by the intensity profile, ${\cal I}(\bvec{y})$.
For a circular source with
dimensionless radius $R_{\rm s} = R_{\rm s}/\eta_0$, and a
uniform intensity profile we have
(eg. Pei \shortcite{PEI93},  or Schneider et al. \shortcite{SCHNEIDER92} 
for an alternative formulation)
\begin{equation}
\mu_{\rm e}(y) = \frac{1}{\pi R_{\rm s}^2} \int_{\vert y - R_{\rm s} \vert}^{y + R_{\rm s}}
\!\!\!\!\!\!{\rm d}t \frac{\sqrt{t^2 -4}\left(R_{\rm s}^2 - y^2 + t^2\right)}
{\sqrt{R_{\rm s}^2 - (y-t)^2}\sqrt{(y+t)^2 - R_{\rm s}^2}}.
\label{eqn:analytic}
\end{equation}
\eqn{analytic} approaches the point source solution as \hbox{$y \rightarrow
\infty$} or $R_{\rm s} \rightarrow 0$, and the maximum magnification is
\begin{equation}
\mu_{\rm e, max} = \frac{\sqrt{R_{\rm s}^2 + 4}}{R_{\rm s}}
\label{eqn:muemax}
\end{equation}
at $y= 0$.

\bsp 

\label{lastpage}

\end{document}